\documentclass[sigconf,nonacm,natbib=false]{acmart}

\usepackage{booktabs}
\usepackage{adjustbox}
\usepackage{caption}
\usepackage{multirow}
\usepackage{dashrule}
\usepackage{makecell}
\usepackage[most]{tcolorbox}
\usepackage{lmodern} 
\usepackage{courier} 
\usepackage{ragged2e}
\usepackage{graphicx}
\usepackage{subcaption}
\usepackage{arydshln}  
\usepackage{array}
\usepackage{makecell}
\usepackage{url}
\usepackage{hyperref}

\usepackage{hyperref} 
\usepackage{footnote} 
\usepackage{tablefootnote}

\usepackage{graphicx} 
\usepackage{array}

\usepackage{tikz}
\usetikzlibrary{trees}

\usepackage{etoolbox} 

\newcommand{\mycomment}[1]{}

\tcbset{
  mypromptbox/.style={
    colback=white,
    colframe=gray!60,
    boxrule=0.5pt,
    arc=3pt,
    fontupper=\small\ttfamily,
    left=4pt, right=4pt, top=4pt, bottom=4pt,
  }
}

\setcopyright{acmlicensed}
\copyrightyear{2025}
\acmYear{2025}
\acmDOI{XXXXXXX.XXXXXXX}

\RequirePackage[
  datamodel=acmdatamodel,
  style=acmnumeric,
  ]{biblatex}

\begin{document}

\title{Ontology-Guided Query Expansion for Biomedical Document Retrieval using Large Language Models}


\author{Zabir Al Nazi}
\affiliation{%
 \institution{University of California Riverside}
 \city{Riverside}
 \state{California}
 \country{USA}}
\email{znazi002@ucr.edu}

\author{Vagelis Hristidis}
\affiliation{%
 \institution{University of California Riverside}
 \city{Riverside}
 \state{California}
 \country{USA}}
\email{vagelis@cs.ucr.edu}

\author{Aaron Lawson Mclean
}
\affiliation{%
 \institution{Friedrich Schiller University Jena}
 \city{Jena}
 \country{Germany}}
\email{aaron.lawsonmclean@med.uni-jena.de}

\author{Jannat Ara Meem}
\affiliation{%
 \institution{University of California Riverside}
 \city{Riverside}
 \state{California}
 \country{USA}}
\email{jmeem001@ucr.edu}

\author{Md Taukir Chowdhury}
\affiliation{%
 \institution{University of California Riverside}
 \city{Riverside}
 \state{California}
 \country{USA}}
\email{mchow068@ucr.edu}

\mycomment{
\begingroup
\small  
\author{Zabir Al Nazi}
\affiliation{University of California Riverside
Riverside, CA, USA}
\email{znazi002@ucr.edu}

\author{Vagelis Hristidis}
\affiliation{University of California Riverside
Riverside, CA, USA}
\email{vagelis@cs.ucr.edu}

\author{Aaron Lawson Mclean}
\affiliation{Friedrich Schiller University Jena
Jena, Germany}
\email{aaron.lawsonmclean@med.uni-jena.de}

\author{Jannat Ara Meem}
\affiliation{University of California Riverside
Riverside, CA, USA}
\email{jmeem001@ucr.edu}

\author{Md Taukir Chowdhury}
\affiliation{University of California Riverside
Riverside, CA, USA}
\email{mchow068@ucr.edu}
\endgroup
}

\renewcommand{\shortauthors}{}


\begin{abstract}

Effective Question Answering (QA) on large biomedical document collections requires effective document retrieval techniques. The latter remains a challenging task due to the domain-specific vocabulary and semantic ambiguity in user queries. We propose \textit{BMQExpander}, a novel ontology-aware query expansion pipeline that combines medical knowledge -- definitions and relationships -- from the UMLS Metathesaurus with the generative capabilities of large language models (LLMs) to enhance retrieval effectiveness.

We implemented several state-of-the-art baselines,  including sparse and dense retrievers, query expansion methods, and biomedical-specific solutions.
We show that \textit{BMQExpander} has superior retrieval performance on three popular biomedical Information Retrieval (IR) benchmarks: NFCorpus, TREC-COVID, and SciFact --  with improvements of up to 22.1\% in NDCG@10 over sparse baselines and up to 6.5\% over the strongest baseline.
Further, \textit{BMQExpander} generalizes robustly under query perturbation settings, in contrast to supervised baselines, achieving up to 15.7\% improvement over the strongest baseline. As a side contribution, we publish our paraphrased benchmarks. 
Finally, our qualitative analysis shows that \textit{BMQExpander} has fewer hallucinations compared to other LLM-based query expansion baselines. 


  
\end{abstract}

\ccsdesc[500]{Information systems~Information retrieval}
\ccsdesc[300]{Information systems~Retrieval models and ranking}
\ccsdesc[300]{Information systems~Query representation and understanding}
\ccsdesc[300]{Information systems~Query reformulation}
\ccsdesc[300]{Information systems~Ontologies}
\ccsdesc[100]{Information systems~Learning to rank}
\ccsdesc[100]{Information systems~Neural networks}

\keywords{Biomedical Information Retrieval, Medical Ontology, Query Expansion, UMLS Metathesaurus, Large Language Models, Chain of Thought.}


\maketitle

\section{Introduction}

Biomedical document retrieval is a critical component in Question Answering (QA) and conversation QA pipelines \cite{qu2020open}. It is essential for doctors and researchers to make clinical decision-making and stay up-to-date with the latest biomedical knowledge and clinical guidelines \cite{sivarajkumar2024clinical}.


However, retrieving relevant documents in response to a biomedical user query is challenging due to the complexity and potential ambiguity of biomedical language and terminology. Queries are often short, ambiguous, or lack precise terminology, which limits the effectiveness of traditional retrieval models \cite{almasri2016comparison, malik2022hybrid, xu2024bmretriever}. Query expansion techniques have the potential to mitigate these issues by augmenting the initial query with semantically related terms or phrases, thereby increasing recall and improving alignment to the user’s intent. 


However, classical expansion techniques often rely on exact matching and struggle with the semantic complexity of biomedical language \cite{tamine2021semantic}. To address these semantic limitations, recent work has turned to LLM-based query expansion \cite{wang2023query2doc,jagerman2023query, chen2024analyze}, which mainly employs LLM prompting and relies on the generalization capabilities of LLMs to generate additional context or reformulated queries.  
Unfortunately, LLMs can generate plausible-sounding but inaccurate or fabricated biomedical terms during query expansion. These incorrect expansions may shift the original intent of the query or introduce clinically invalid terminology. Hence, LLM-based query expansions can sometimes introduce hallucinated or inaccurate information, especially when not grounded in trusted biomedical knowledge sources. This is a significant concern in medical information retrieval, where factual accuracy is essential. 
Moreover, generating expansions using LLMs can be resource-intensive, particularly when incorporating advanced prompting techniques such as iterative refinement, or multi-stage pipelines. For example, retrieving top-ranked documents via a baseline retriever and then using them to condition the LLM adds additional computational steps \cite{chen2024analyze, lei-etal-2024-csqe}. 

These issues raise an important research question: \emph{Can we design a biomedical query expansion framework that reduces dependence on LLM generation and instead leverages trusted, structured medical knowledge for generating context to guide LLMs toward more accurate and factually grounded query expansions?}

In this paper, to address the aforementioned limitations, we propose \textbf{BMQExpander}, a pipeline that integrates structured biomedical ontologies with the generative abilities of LLMs in a controlled way. Our approach is \textbf{LLM-agnostic} - the expansion is primarily driven by medical definitions and concept relationships retrieved from the \textbf{UMLS Metathesaurus} \cite{bodenreider2004unified}. 
Our method directly addresses the problem of hallucinated or clinically invalid expansions by grounding the generation of LLM in trusted biomedical ontologies. This is supported by existing work \cite{munnangi2024fly} that suggests that incorporating precise definitions and reliable external knowledge into the prompt context enhances LLM understanding, improves reasoning quality, and reduces the likelihood of factual errors.


In BMQExpander, we extract key medical terms from a query using an LLM with in-context examples, map them to UMLS concepts, and retrieve curated definitions and semantic relationships from biomedical vocabularies (e.g., MeSH, SNOMED, NCI) \cite{lipscomb2000medical, donnelly2006snomed, de2023national}. We represent the relevant concept hierarchies and associations as a pruned semantic graph by heuristically selecting domain-specific edges. We prompt an LLM with the serialized graph and generate a medically grounded, context-rich pseudo-document as a response to the original query. The query is then expanded with this output and used for retrieval via a standard ranking function.

We implemented three categories of baseline retrievers: sparse retrieval using BM25, dense retrieval with both general-purpose and biomedical-specific models, and query expansion approaches that include both traditional medical methods and recent LLM-based techniques. We experimented on  three popular biomedical Information Retrieval (IR) benchmarks: NFCorpus, TREC-COVID, and SciFact \cite{boteva2016full, voorhees2021trec, wadden-etal-2020-fact}.
Our experiments show that \textbf{BMQExpander} outperforms the sparse baseline by up to 22.1\% in NDCG@10.
Experimentally we also show that \textbf{BMQExpander} is robust with respect to the choice of LLMs and prompting strategies. 
Our datasets, results and code are published online\footnote{\url{https://github.com/zabir-nabil/ontology-guided-query-expansion}}.

To summarize, the main contributions of this paper are:
\begin{itemize}
    \item We propose \textit{BMQExpander}, a novel query expansion framework for the biomedical domain that uses medical knowledge from UMLS to generate grounded and context-rich expansions. 


    
    \item We perform extensive experiments on three benchmark biomedical IR datasets, comparing to a wide range of baselines, including query expansion methods, LLM prompting approaches, and both dense and sparse retrieval models.

    \item We created    three new perturbed biomedical retrieval datasets to assess the generalization capabilities of existing biomedical retrieval methods. We found that these methods significantly degrade for such new datasets, in contrast to BMQExpander which generalizes to unknown datasets. 
    
    \item We perform a qualitative analysis with the help of a medical domain expert, which shows that \textit{BMQExpander} has fewer hallucinations and higher medical accuracy.

\end{itemize}

The remainder of the paper is organized as follows. Section 2 reviews related work. Section 3 details our proposed methodology. Section 4 introduces our dataset contribution. Section 5 describes the experimental setup and presents the results. Finally, Section 6 concludes the paper.

\section{Related Work}


We begin by reviewing sparse retrieval approaches, followed by dense retrieval models. Next, we cover query expansion techniques, including PRF and LLM-based methods. For each of these categories, we also include a focused discussion on their applications and adaptations in biomedical document retrieval.

\subsection{Sparse Retrieval}
Sparse retrieval methods, such as BM25, TF-IDF, and Vector Space Model have been widely used in traditional information retrieval systems due to their simplicity, efficiency, and strong performance in lexical matching tasks. These models rely on exact term overlap and use inverted indexes to retrieve documents efficiently. Despite their success, sparse retrievers often struggle with vocabulary mismatch and limited semantic understanding, prompting efforts to enhance them through techniques such as learning-to-rank and query expansion \cite{hambarde2023information}.

However, recent advances have introduced neural-enhanced sparse retrieval methods that either re-weight traditional terms using deep learning models \cite{dai2019context} or learn sparse representations directly in latent space \cite{formal2021splade}. These approaches retain the indexing efficiency of symbolic models while incorporating semantic understanding, bridging the gap between classical term-based retrieval and semantic matching.

In the \textbf{biomedical domain}, recent methods have improved sparse retrievers by addressing domain shift and vocabulary mismatch using unsupervised adaptation. Notably, \cite{vast2024simple} adapt SPLADE \cite{formal2021splade} with domain-specific pretraining, while \cite{nishida2023sparse} introduce SPARC, which enhances retrieval via domain-tuned embeddings and learnable sparsity for efficient biomedical question answering.

\subsection{Dense Retrieval Models}

Dense retrieval methods map queries and documents into a shared vector space using neural network encoders, enabling semantic similarity matching beyond the exact token overlap. Unlike sparse retrievers like BM25, dense retrievers learn continuous representations that better capture semantic meaning. Dense Passage Retrieval (DPR) \cite{karpukhin2020dense} introduced a dual-encoder architecture using BERT, and has led to many follow-up studies on dense retrieval.

Recent surveys \cite{zhao2024dense} have reviewed the progress in dense retrieval, including multi-vector models and hybrid dense-sparse approaches. While dense models often provide stronger semantic matching, they also require significant training data and computational resources, motivating ongoing work on efficiency and scalability.

In the \textbf{biomedical domain}, dense retrievers like MedCPT \cite{jin2023medcpt} and BMRetriever \cite{xu2024bmretriever} have been developed to better handle domain-specific terminology and complex semantics. These models demonstrate strong performance on biomedical benchmarks by leveraging pre-training in medical corpora and adapting retrieval strategies to clinical language.

\subsection{Query Expansion}

Query expansion (QE) aims to enhance the original user query by incorporating additional terms that are semantically related, thereby improving retrieval performance. Classical QE techniques utilize lexical resources such as thesauri or ontologies to find synonyms and related words for query terms \cite{bhogal2007review}. 

Global analysis methods, for instance, automatically expand each query term with its synonyms and related terms. Stemming can also be used to include various morphological forms of words, further increasing recall \cite{azad2019query}.

More recently, neural-network based approaches to query expansion have gained traction. Techniques based on word embeddings identify semantically related terms by measuring vector similarity in embedding space. Contextualized models like BERT-QE \cite{zheng2020bert} use transformer-based embeddings to select relevant document chunks for expansion and have shown significant gains in effectiveness.

\subsubsection{Pseudo-Relevance Feedback}

Pseudo-relevance feedback (PRF) assumes that the top-ranked documents from an initial retrieval step are relevant and uses them to refine the query. Expansion terms are extracted from these documents and added to the original query to bridge the lexical gap between queries and documents. PRF has been widely adopted and is known to improve retrieval performance in both classical and neural settings.

Recent advances in PRF have focused on combining it with dense representations. ColBERT-PRF \cite{wang2023colbert} applies PRF in the context of late-interaction dense retrievers. Similarly, BERT-QE \cite{zheng2020bert} integrates PRF by selecting semantically meaningful document spans, leveraging contextual representations for expansion.

\subsubsection{Query Expansion using Large Language Model} LLM-based query expansion leverages large language models to generate semantically rich expansions that address limitations of keyword-based expansion. Unlike traditional methods, LLMs can incorporate contextual understanding and domain knowledge to produce semantically relevant queries. Recently proposed, the Query2doc \cite{wang2023query2doc} method enhances information retrieval by generating pseudo-documents through few-shot prompting of large language models (LLMs), which are then used to expand the original query. Similarly, the AGR framework uses a three-step prompting strategy: Analyze, Generate, and Refine to create answer-focused expansions for answering zero-shot questions, achieving strong performance in and out of the domain \cite{chen2024analyze}. Corpus-Steered Query Expansion (CSQE) grounds expansions in corpus content by extracting key sentences from initially retrieved documents, helping reduce hallucinations and outdated information \cite{lei-etal-2024-csqe}.

\subsubsection{Biomedical Query Expansion} In the biomedical domain, hybrid approaches that combine clinical knowledge with dense representations have shown promise in addressing vocabulary mismatch and domain specificity. \cite{malik2022hybrid} proposed a framework that integrates clinical diagnosis information with domain-specific and domain-agnostic word embeddings to expand queries using both symbolic and semantic signals. Similarly, \cite{ojha2021metadata} introduced a metadata-driven semantically aware expansion method that aligns PubMed ontologies with enriched query concepts derived from external knowledge bases and semantic similarity metrics. These domain-adapted techniques demonstrate the importance of grounding query expansion in biomedical knowledge for improved document retrieval.




\mycomment{
\paragraph{Sparse Retrieval}

\paragraph{Dense Retrieval}

\paragraoh{LLM-based Retrieval methods}

\paragraph{Query Expansion}  include PRF here

\paragraph{Biomedical Document Retrieval} here focuses on biomedical methods from all of the above categories

Retrieving relevant documents in remains a long-standing challenge in information retrieval \cite{crouch2002improving, marrero2012information}. Sparse retrieval methods, such as BM25, are widely used due to their simplicity and effectiveness, but often struggle with lexical mismatch between query terms and document content \cite{lin2021pyserini}. To address this, a large body of work has focused on improving the query representation through query expansion and refinement, where additional semantically related terms are added to improve retrieval performance. Classical methods such as pseudo-relevance feedback (PRF) rely on top-ranked documents to extract expansion terms \cite{almasri2016comparison}, while more recent work leverages neural networks and large language models (LLMs) for generating richer expansions \cite{imani2019deep, wang2023query2doc, jagerman2023query, chen2024analyze}.

In this section, we review three main lines of research that are most relevant to our work: \textbf{query expansion} methods that aim to enrich the input query with semantically related terms or content \cite{wang2023query2doc}, \textbf{dense retrieval models} that learn vector representations to enable semantic matching between queries and documents \cite{xu2024bmretriever}, and \textbf{pseudo-relevance feedback} techniques that expand queries using terms extracted from the top retrieved documents \cite{almasri2016comparison, chen2024analyze}.

}

\begin{figure*}[t]
    \centering
    \includegraphics[width=0.95\textwidth, trim=0 40 0 0, clip]{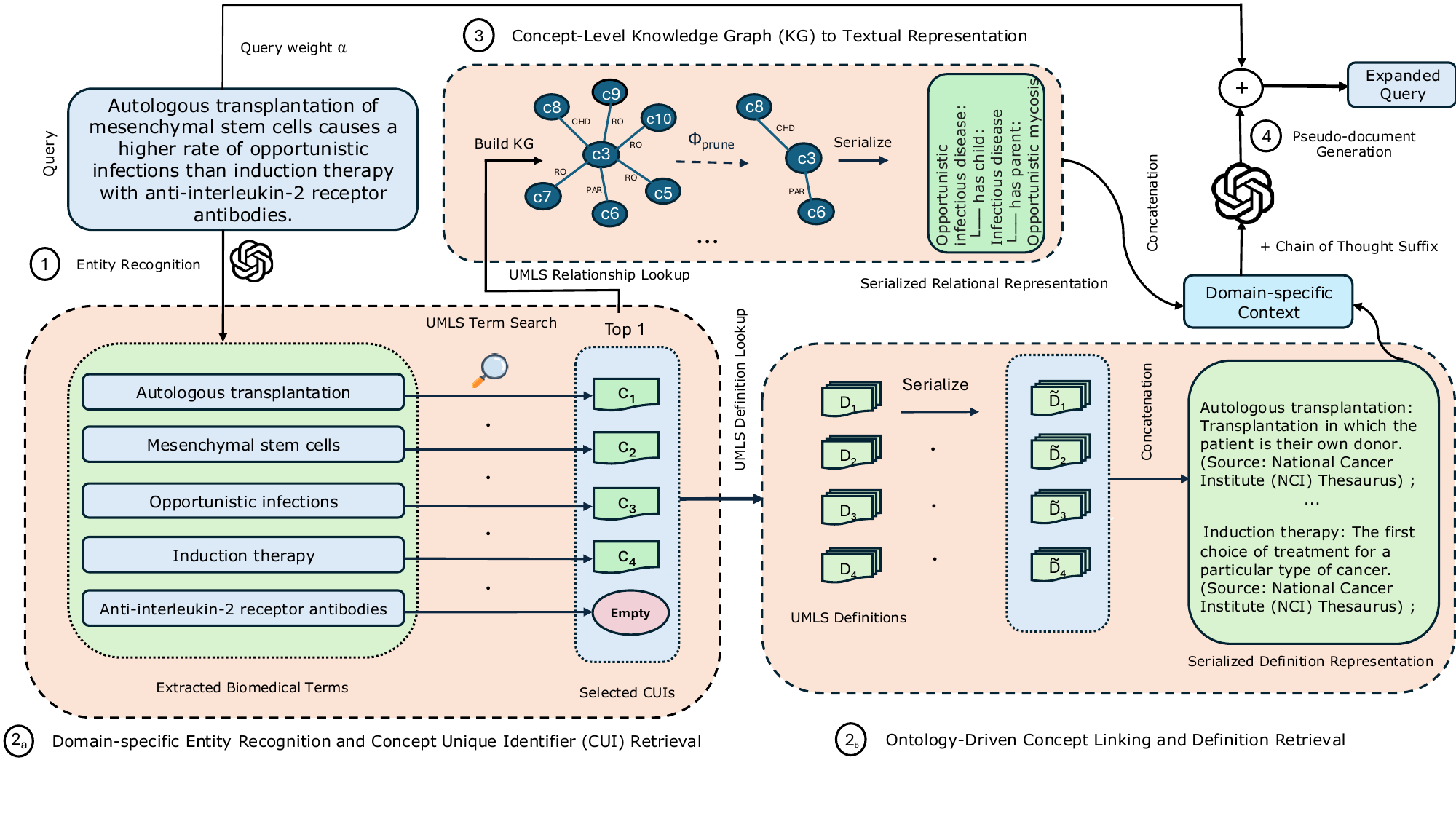}
    \caption{Schematic diagram of the proposed method. Here, $c3$ represents CUI C0029118, identifier of "Opportunistic Infections".} 
    \label{fig:proposed_method}
\end{figure*}

\section{Methodology}

Let $\mathcal{Q}$ be the space of user queries and $\mathcal{D}$ the corpus of biomedical documents. Given a query $q \in \mathcal{Q}$, our objective is to construct a medically enriched, semantically faithful expanded query $q' \in \mathcal{Q}'$ where $\mathcal{Q}'$ is the space of expanded queries such that retrieval with $q'$ yields improved clinical retrieval performance and robustness compared to using $q$ directly.


We propose a five-stage pipeline, shown in Figure~\ref{fig:proposed_method}:

\begin{enumerate}
    \item Domain-Specific Entity Recognition
    \item Ontology-Driven Concept Linking and Definition Retrieval
    \item Concept-Level Knowledge Graph Construction
    \item Ontology-Guided Pseudo-document Generation 
    \item Weighted Query Expansion and Retrieval
\end{enumerate}

\subsection{Domain-Specific Entity Recognition}

We begin by identifying key biomedical terms in a user query $q \in \mathcal{Q}$. Let $\mathcal{V}$ denote the vocabulary of biomedical concepts. We define an entity recognition function, $\mathcal{T} : \mathcal{Q} \rightarrow 2^{\mathcal{V}}$ that maps a query to a subset of biomedical terms $T_q$.

\[
T_q = \mathcal{T}\left(q\right) \subseteq \mathcal{V}, \quad T_q = \{ t_1, t_2, \ldots, t_n \}
\]

We instantiated $\mathcal{T}$ using few-shot prompting of an LLM with in-context examples to extract key biomedical terms that are central to the query, including complex terms that may be misinterpreted or overlooked by the model without formal definitions and structured context. Our approach avoids trivial or shallow term selection, generating conceptually informative terms that improve downstream query expansion. If the model determines that no medically meaningful terms are present, we assign $T_q = \emptyset$.

We use the following prompt to extract relevant medical terms from the query.

\begin{tcolorbox}[
  colback=purple!5,  
  colframe=gray!80!black,
  boxrule=0.4pt,
  arc=2pt,
  enhanced,
  title=Domain-Specific Entity Recognition Prompt,
  left=4pt, right=4pt, top=1pt, bottom=1pt,
]

\begin{tcolorbox}[
  colback=white,
  colframe=purple!40,
  boxrule=0.2pt,
  sharp corners,
  left=4pt, right=4pt, top=2pt, bottom=2pt,
  title=\textbf{\sffamily System Prompt},
  fonttitle=\bfseries\sffamily\scriptsize,
  fontupper=\ttfamily\footnotesize,
  enhanced
]
You are a biomedical information retrieval assistant.
\end{tcolorbox}


\begin{tcolorbox}[
  colback=white,
  colframe=purple!40,
  boxrule=0.2pt,
  sharp corners,
  left=4pt, right=4pt, top=1pt, bottom=1pt,
  title=\textbf{\sffamily User Prompt},
  fonttitle=\bfseries\sffamily\scriptsize,
  fontupper=\ttfamily\footnotesize,
  enhanced
]
Your task: Extract key medical terms from the query. If the query lacks significant medical terms, return an empty list.
\end{tcolorbox}


\begin{tcolorbox}[
  colback=white,
  colframe=purple!40,
  boxrule=0.2pt,
  sharp corners,
  left=4pt, right=4pt, top=1pt, bottom=1pt,
  title=\textbf{\sffamily In-Context Examples},
  fonttitle=\bfseries\sffamily\scriptsize,
  fontupper=\ttfamily\footnotesize,
  enhanced
]
Query: Dietary Treatment of Crohn’s Disease \\
Terms: [Dietary Treatment, Crohn’s Disease] \\[4pt]
\ldots
\end{tcolorbox}


\begin{tcolorbox}[
  colback=white,
  colframe=purple!40,
  boxrule=0.2pt,
  sharp corners,
  left=4pt, right=4pt, top=1pt, bottom=1pt,
  title=\textbf{\sffamily Input Query},
  fonttitle=\bfseries\sffamily\scriptsize,
  fontupper=\ttfamily\footnotesize,
  enhanced
]
Query: \fbox{\texttt{\textbf{\{query\}}}} 

Terms: [\ldots]
\end{tcolorbox}

\end{tcolorbox}

\subsection{Ontology-Driven Concept Linking and Definition Retrieval}

\subsubsection{Concept Linking via UMLS Search}

Following biomedical term extraction, we link each identified term $t_i \in T_q$ to its corresponding Concept Unique Identifier (CUI) $c_i$ using the UMLS Metathesaurus. Let $\mathcal{CUI}$ denote the space of CUIs. We define a concept normalization function that maps each biomedical term to a corresponding CUI, $
\mathcal{C} : \mathcal{V} \rightarrow \mathcal{CUI}$.

 We then define the set of CUIs for a given query as,

\[
C_q = \{ \mathcal{C}\left(t_i\right) \mid t_i \in T_q \} = \{ c_1, c_2, \ldots, c_n \}
\]

To resolve each term $t_i \in T_q$ to its corresponding CUI $c_i \in \mathcal{CUI}$, we query the UMLS Metathesaurus using an exact-match search strategy. Specifically, we use the UMLS API\footnote{\url{https://documentation.uts.nlm.nih.gov/rest/home.html}} to perform concept lookups to return the concept that exactly matches the lexical form of $t_i$ to avoid the noise and ambiguity often introduced by partial or fuzzy matching.

If the exact-match search returns multiple CUIs for a term, we select the top concept and retrieve its identifier. Terms with no exact match are discarded to prevent erroneous mappings and preserve semantic precision. The resulting set of CUIs, $C_q$, provides a structured representation of the query, accurately aligning biomedical ontologies in subsequent retrieval stages.

\subsubsection{Definition Retrieval}

Each CUI $c_i \in C_q$ is associated with one or more definitions retrieved from UMLS. We source the definitions from a set of curated biomedical vocabularies,
\[
\mathbb{B} = \{ \text{MSH}, \text{SNOMEDCT\_US}, \text{NCI}, \text{CSP} \}
\]

Here, \texttt{MSH} denotes the \textit{Medical Subject Headings (MeSH)} \cite{lipscomb2000medical}, \texttt{SNOMEDCT\_US} indicates the \textit{Systematized Nomenclature of Medicine Clinical Terms, US Edition} \cite{donnelly2006snomed}, \texttt{NCI} refers to the \textit{National Cancer Institute Thesaurus} \cite{de2023national}, and \texttt{CSP} corresponds to the \textit{CRISP Thesaurus} \cite{national2002computer}. We restrict definition retrieval to these four vocabularies, as they consistently provide concise, well-structured definitions and offer better coverage across our evaluation datasets which is supported by previous work \cite{zivaljevic2022utility}.

For each concept $c_i \in C_q$, we retrieve definitions from 
biomedical vocabularies. Each definition is paired with its source citation to maintain semantic transparency. Formally, we define the definition set for a concept $c_i$ as,

\[
D_i = \left\{ (d_{ij}, b_{ij}) \;\middle|\; d_{ij} \in \mathbb{D},\; b_{ij} \in \mathbb{B},\; j = 1, \ldots, k_i \right\}
\]

where $d_{ij}$ is a natural language definition and $b_{ij}$ is the source vocabulary (e.g., \texttt{MSH}, \texttt{NCI}, \texttt{SNOMEDCT\_US}, \texttt{CSP}) from which it was retrieved.

To ensure clarity and consistency during downstream prompting, we serialize all definitions for a given concept $c_i$ into a single structured string, $\tilde{D_i}$.

Finally, \( \bar{D}_q \) is the concatenation of the serialized definition sets \( \tilde{D_i} \) for all concepts \( c_i \in C_q \), representing the full definition context for query \( q \).

This format allows multiple definitions to be combined in a readable way while preserving the source of each. By linking each concept to trusted biomedical vocabularies, it is expected to help reduce ambiguity and mitigate hallucination in downstream generation \cite{huang2023citation, byun2024reference}. The standardized schema is exemplified below.

\label{app:serialization}
\begin{tcolorbox}[
  colback=red!5,  
  colframe=gray!80!black,  
  boxrule=0.4pt,
  arc=2pt,
  enhanced,
  title=Serialization of Definitions,
  left=4pt, right=4pt, top=4pt, bottom=4pt,
]
\fbox{\textit{\texttt{\textbf{Concept Name}}}}: definition$_1$ (Source: Vocabulary$_1$); definition$_2$ (Source: Vocabulary$_2$); $\ldots$ definition$_k$ (Source: Vocabulary$_k$);

\vspace{4pt}

\begin{tcolorbox}[
  colback=white,
  colframe=red!40,
  boxrule=0.2pt,
  sharp corners,
  left=4pt, right=4pt, top=2pt, bottom=2pt,
  title=\textbf{\sffamily Example},
  fonttitle=\bfseries\sffamily\scriptsize,
  fontupper=\ttfamily\footnotesize,
  enhanced
]
\texttt{\textbf{Lymphatic Filariasis}}: A clinical disorder that is caused by obstruction of the lymphatic system years after filarial infection. It is characterized by painful and profound lymphedema, resulting in significant swelling (elephantiasis) of extremities and genitals. (Source: National Cancer Institute (NCI) Thesaurus); Parasitic infestation of the human lymphatic system by WUCHERERIA BANCROFTI or BRUGIA MALAYI. It is also called lymphatic filariasis. (Source: MeSH);
\end{tcolorbox}
\end{tcolorbox}

\subsection{Concept-Level Knowledge Graph Construction}

To enrich each query with structured biomedical knowledge, we construct a local semantic graph around each extracted concept. 
For each concept $c_i \in C_q$, we define a directed labeled graph $G_i = (V_i, E_i)$, where the node set $V_i$ includes $c_i$ and its neighboring CUIs retrieved via the UMLS Metathesaurus, and each edge $(c_i, c_j, r) \in E_i$ represents a semantic relation $r \in \mathcal{R}$ between $c_i$ and $c_j$.

Since UMLS contains a large and heterogeneous set of relation types - many of which are overly broad or semantically redundant for biomedical retrieval - we prune the relation set using a filtering function, $\phi_{\text{prune}} : \mathcal{R} \rightarrow \{0,1\}$
that retains only medically meaningful relations:
\[
\mathcal{R}' = \left\{ r \in \mathcal{R} \;\middle|\; \phi_{\text{prune}}(r) = 1 \right\}
\]
\[
= \{ \texttt{CHD},\ \texttt{PAR},\ \texttt{SY},\ \texttt{RO},\ \texttt{RO:has\_associated\_morphology} \}
\]

\texttt{CHD} (has child) and \texttt{PAR} (has parent) capture hierarchical \textit{is-a} relationships useful for modeling taxonomic structures; \texttt{SY} (synonym) links conceptually equivalent terms, aiding in synonym-aware retrieval; \texttt{RO} (related other) expresses general associative relationships between two concepts; and \texttt{RO:has\_associated\_morphology} is a specific sub-relation indicating morphological associations (e.g., linking a disease to its pathological form).

Applying this filter, we obtain a pruned edge set for each concept:
\[
E_i' = \left\{ (c_i, c_j, r) \in E_i \;\middle|\; r \in \mathcal{R}' \right\}, \quad \text{yielding} \quad G_i' = (V_i, E_i')
\]

This pruned graph removes weak or noisy edges and ensures an interpretable and reliable concept-level graph.

We serialize the structured pruned graph $G_i'$  into a textual representation $s_i \in \mathcal{S}_{\text{struct}}$, where $\mathcal{S}_{\text{struct}}$ denotes the space of structured textual encodings produced via serialization, to provide as input to the LLM. Each node $c_i \in V_i$ is expressed as a hierarchy of labeled edges, formatted in an indented, human-readable layout, with an example provided below.

\label{app:serializationrel}
\begin{tcolorbox}[
  colback=green!5, 
  colframe=gray!80!black, 
  arc=2pt, 
  boxrule=0.3pt, 
  title=Serialization of Relationships, 
  fontupper=\ttfamily\small
]
\textbf{Carcinoma of breast}: \\
\hspace*{1em} $\llcorner$ has parent: Infiltrating duct carcinoma \\
\hspace*{1em} $\llcorner$ is synonymous with: Breast cancer
\end{tcolorbox}

For a given query $q$, we aggregate the concept-level relational contexts to form a unified prompt input. Let $S_q = \{s_1, s_2, \ldots, s_n\}$ be the set of serialized representations for each pruned concept-level graph $G_i'$, where $s_i \in \mathcal{S}_{\text{struct}}$ corresponds to concept $c_i \in C_q$. We define the full relational context string as, $\bar{S}_q = \texttt{concat}\left(s_1, s_2, \ldots, s_n\right)$
where \( \bar{S}_q \in \mathcal{S}_{\text{struct}}^* \) denotes the concatenated structured relational context used to guide the language model during generation.

This textual format retains the semantic integrity of the ontology while making the relational structure directly usable in language model prompts. By encoding only high-confidence, ontology-derived links, it reduces ambiguity and helps prevent hallucinations during generation. This ensures that the expanded queries remain medically grounded and aligned with the intended biomedical context.

\subsection{Contextualized Generation Guided by Biomedical Ontology}

To generate a semantically enriched pseudo-document, we prompt the language model with the original query \( q \), the serialized definition context \( \bar{D}_q \), and the serialized relational context \( \bar{S}_q \), when available from prior stages of the pipeline. The resulting pseudo-document is defined as:
\[
p_q = \mathcal{G}\left(q,\ \bar{D}_q,\ \bar{S}_q\right)
\]
where \( p_q \in \Sigma^* \) is a natural language string used as an expanded query for downstream retrieval. The function \( \mathcal{G} \) represents LLM-based generation guided by biomedical definitions and ontology-derived relationships. 

The model is prompted with a maximum generation limit of 512 tokens. Furthermore, we append a chain-of-thought suffix, \textbf{“Give the rationale before answering”} to encourage step-by-step reasoning in the generated output.

The prompt is formatted as,

\begin{tcolorbox}[
  colback=blue!5,
  colframe=gray!80!black,
  arc=2pt,
  boxrule=0.3pt,
  title=Ontology-Guided Pseudo-document Prompt,
  fontupper=\ttfamily\small]

Given a query, relevant medical definitions and relationships; write an answer to the query.

\vspace{3pt}

Query: \fbox{\texttt{\textbf{\{query\}}}}

\vspace{3pt}

Definitions: \fbox{\texttt{\textbf{\{definitions\}}}}

\vspace{3pt}

\vspace{3pt}
Relationships: \fbox{\texttt{\textbf{\{relationships\}}}}

\vspace{2pt}

\end{tcolorbox}


\begin{table*}[ht]
    \centering
    \caption{Dataset statistics and query samples}
    \renewcommand{\arraystretch}{1.2} 
    \resizebox{\textwidth}{!}{%
    \begin{tabular}{l | c | c | c | >{\centering\arraybackslash}m{1.5cm} | >{\centering\arraybackslash}m{1.8cm} | c | >{\centering\arraybackslash}m{5.5cm} | >{\centering\arraybackslash}m{5.8cm}}

        \hline
        \textbf{Dataset} 
        & \textbf{\# Docs} 
        & \textbf{\# Queries} 
        & \textbf{Avg. Doc Length} 
        & \multicolumn{2}{c|}{\textbf{Avg. Query Length}} 
        & \textbf{\# Unique Words} 
        & \textbf{Original Query Example} 
        & \textbf{Paraphrased Query Example} \\
        \cline{5-6}
        & & & & \textbf{Original} & \textbf{Paraphrased} & & & \\
        \hline

        \multirow{3}{*}{NFCorpus-P} 
        & \multirow{3}{*}{3,633} 
        & \multirow{3}{*}{323} 
        & \multirow{3}{*}{219.656} 
        & \multirow{3}{*}{3.272} 
        & \multirow{3}{*}{5.344} 
        & \multirow{3}{*}{36,735} 
        & Cancer Risk From CT Scan Radiation 
        & Risk of Developing Cancer Due to Radiation from CT Scans \\
        \cline{8-9}
        & & & & & & 
        & BPH 
        & What is benign prostatic hyperplasia? \\
        \cline{8-9}
        & & & & & & 
        & Preventing Brain Loss with B Vitamins? 
        & Can B vitamins help stop brain deterioration? \\
        \hline

        \multirow{3}{*}{TREC-COVID-P} 
        & \multirow{3}{*}{1,71,332} 
        & \multirow{3}{*}{50} 
        & \multirow{3}{*}{148.192} 
        & \multirow{3}{*}{10.600} 
        & \multirow{3}{*}{11.860} 
        & \multirow{3}{*}{3,83,990} 
        & what causes death from Covid-19? 
        & What factors lead to fatalities in Covid-19 cases? \\
        \cline{8-9}
        & & & & & & 
        & What is the mechanism of cytokine storm syndrome on the COVID-19? 
        & How does a cytokine storm occur in COVID-19 patients? \\
        \cline{8-9}
        & & & & & & 
        & Does Vitamin D impact COVID-19 prevention and treatment? 
        & Can Vitamin D affect the prevention and treatment of COVID-19? \\
        \hline

        \multirow{3}{*}{SciFact-P} 
        & \multirow{3}{*}{5,183} 
        & \multirow{3}{*}{300} 
        & \multirow{3}{*}{200.811} 
        & \multirow{3}{*}{12.493} 
        & \multirow{3}{*}{14.630} 
        & \multirow{3}{*}{50,571} 
        & DMRT1 is a sex-determining gene that is epigenetically regulated by the MHM region. 
        & The MHM region regulates the DMRT1 gene, which plays a role in sex determination, through epigenetic mechanisms. \\
        \cline{8-9}
        & & & & & & 
        & Birth-weight is positively associated with breast cancer. 
        & Higher birth-weight is linked to an increased risk of breast cancer. \\
        \cline{8-9}
        & & & & & & 
        & High levels of copeptin decrease risk of diabetes. 
        & Elevated copeptin levels lower the risk of developing diabetes. \\
        \hline
    \end{tabular}
    }
    
    \label{tab:dataset_stats_queries}
\end{table*}

\subsection{Weighted Query Expansion and Retrieval}

To form the final expanded query, we combine the original query $q \in \mathcal{Q}$ with the pseudo-document $p_q$ generated by the language model. We define the expanded query $q' \in \mathcal{Q}'$ as,

\[
q' = \underbrace{q \oplus q \oplus \cdots \oplus q}_{\alpha\ \text{times}} \oplus\ p_q
\]

where $\oplus$ denotes string concatenation, and the weighting factor $\alpha$ controls the relative influence of the original query versus the generated content. In our main experiments, we use $\alpha = 5$ as commonly used in prior work \cite{wang2023query2doc, jagerman2023query}, where it has been demonstrated to reliably balance the preservation of the original query intent with the incorporation of generated expansion content, without the need for dataset-specific tuning.

We use the BM25 retrieval model to rank documents in the corpus $\mathcal{D}$ based on the expanded query $q'$. The pseudo-document \( p_q \) serves as a generator of likely document-side keywords, helping bridge lexical gaps between the query and relevant documents. The original query is repeated with a weighting factor \( \alpha \) to ensure that the retrieval remains anchored to the user's original intent and is not overwhelmed by the generative content.

\mycomment{
\begin{figure}[ht]
    \centering
    \begin{subfigure}[t]{0.48\textwidth}
        \centering
        \includegraphics[width=\textwidth]{kg_before.pdf}
        \caption{Full Knowledge Graph}
        \label{fig:kg-before}
    \end{subfigure}
    \hfill
    \begin{subfigure}[t]{0.48\textwidth}
        \centering
        \includegraphics[width=\textwidth]{pruned_kg.pdf}
        \caption{Pruned Knowledge Graph}
        \label{fig:pruned-kg}
    \end{subfigure}
    
    \caption{Comparison of the full and pruned knowledge graphs.}
    \label{fig:kg-comparison}
\end{figure}
}

\section{Datasets}

We evaluate our method on three widely used benchmarks in biomedical information retrieval: \textbf{NFCorpus}, \textbf{SciFact}, and \textbf{TREC-COVID} \cite{boteva2016full, voorhees2021trec, wadden-etal-2020-fact}.

\textbf{NFCorpus} is a full-text medical information retrieval dataset containing 3,244 natural language queries from \url{NutritionFacts.org}, paired with over 9,000 PubMed-based documents and approximately 170,000 relevance judgments. For evaluation, we follow the BEIR benchmark \cite{thakur2021beir} splits and report results on the held-out test set, which contains 3,633 documents. 

\textbf{SciFact} consists of 1,400 expert-written scientific claims matched with biomedical abstracts, each annotated with supporting or refuting evidence and sentence-level rationales. We use the BEIR test split, comprising 300 queries and their associated documents. 

\textbf{TREC-COVID} is built on the CORD-19 corpus and includes multiple rounds of expert-formulated topics and human-annotated relevance judgments, modeling retrieval for urgent and evolving biomedical questions. Our experiments use the BEIR test split, which includes 50 queries paired with relevance-annotated documents. These datasets vary in query intent, document complexity, and domain specificity, providing a comprehensive evaluation setup.


\mycomment{
\subsection{Benchmark Datasets}
\paragraph{NFCorpus.} 
NFCorpus is a full-text medical information retrieval dataset containing 3,244 natural language queries harvested from NutritionFacts.org. The queries are non-technical and user-oriented, whereas the associated documents—drawn primarily from PubMed—are technical and terminology-heavy. It includes over 9,000 medical documents and nearly 170,000 relevance judgments, with query-document links established via a hierarchy of citation strength. The dataset is split into training (80\%), development (10\%), and testing (10\%) based on query-level partitions.

\paragraph{SciFact.} 
SciFact is a scientific fact-checking benchmark comprising 1,400 expert-written scientific claims paired with biomedical abstracts. Each claim is annotated with supporting or refuting evidence, along with sentence-level rationales. The dataset includes a separate corpus of abstracts with document IDs and metadata. Queries in SciFact are formal claims rather than keyword-style questions, enabling assessment of semantic retrieval capabilities and evidence-grounded reasoning.

\paragraph{TREC-COVID.}
TREC-COVID is a large-scale ad hoc retrieval dataset built on the CORD-19 corpus, a continuously updated collection of scientific articles on COVID-19. Structured over five rounds, the dataset includes hundreds of expert-formulated information needs (topics) and human-annotated relevance judgments. TREC-COVID evaluates the ability of systems to retrieve relevant literature in response to urgent, evolving biomedical questions—reflecting real-world use in dynamic public health crises.
}

\paragraph{Paraphrased Query Benchmarks}

Many information retrieval systems, particularly dense retrievers, have been shown to be sensitive to variations in query phrasing, leading to substantial performance degradation when semantically equivalent queries are expressed differently \cite{penha2022evaluating, arabzadeh2023noisy, cheng2022task}. To systematically assess retrieval robustness under such query perturbations, especially for biomedical domain, we construct paraphrased variants of three standard benchmarks - \textbf{NFCorpus-P}, \textbf{SciFact-P}, and \textbf{TREC-COVID-P} - by generating paraphrased versions of the original queries that preserve their meaning. An overview of the new benchmarks is shown in table \ref{tab:dataset_stats_queries}.

We use \textbf{GPT-4o} to generate paraphrased queries. For each original query, we apply the following prompt:

\begin{tcolorbox}[
  colback=yellow!5,
  colframe=gray!80!black,
  arc=2pt,
  boxrule=0.3pt,
  title=Query Perturbation Prompt,
  fontupper=\ttfamily\small]

\texttt{Paraphrase the following query.} \\
\texttt{Query: \fbox{\textbf{\{query\}}}} \\
\texttt{Paraphrased query:}

\end{tcolorbox}

This approach preserves the original intent and relevance annotations of the datasets while introducing natural lexical and syntactic variation. To ensure quality, all paraphrased queries were manually reviewed by the authors to verify that each reformulation was semantically faithful to the original and free from grammatical or factual errors. These paraphrased benchmarks are used to evaluate the robustness of retrieval systems, allowing us to compare our method against baseline models under semantically equivalent query reformulations. 

\section{Experiments}

We evaluate our proposed method across three benchmark biomedical retrieval datasets: \textbf{NFCorpus}, \textbf{TREC-COVID}, and \textbf{SciFact}. The goal is to retrieve and rank relevant documents given a query. We primarily report \textbf{NDCG@10} as our main evaluation metric for IR. For query perturbation experiment and comparison of LLM backbones, we also report \textbf{mAP@10} and \textbf{Recall@10} to provide a more comprehensive view of performance.

\subsection{Experimental Setup}

We describe the implementation details and baseline methods used for comparison with our proposed approach. Our experiments include comparisons against sparse retrieval, dense retrievers, biomedical-specific retrieval pipelines and recent LLM-based query expansion methods. For the main experiments, we use GPT-4o as the LLM backbone, configured with default parameters. We set $\alpha = 5$ for LLM-based experiments. For ablation study without the LLM, we use a higher value ($\alpha = 50$) to ensure BM25 gives sufficient weight to the original query, since the expansion process in this setting introduces substantially more terms than LLM-based generation, which is limited to 512 tokens. Below, we briefly describe each category of baselines used in our experiments. 

\paragraph{\textbf{Sparse Retriever.}} As a lexical baseline, we use the BM25 retriever implemented via the Pyserini toolkit \cite{lin2021pyserini, thakur2021beir}. BM25 scores documents based on term frequency and inverse document frequency (TF-IDF), and remains a strong and widely adopted sparse retrieval method.

\paragraph{\textbf{Dense Retriever.}}
We evaluate several dense retrieval baselines that encode queries and documents into a shared embedding space using pretrained transformer encoders. These include general-purpose dense retrievers such as Contriever~\cite{izacard2021unsupervised}, GTR-L~\cite{ni2021large}, InstructOR-L~\cite{su2022one}, SGPT~\cite{muennighoff2022sgpt}, SPECTER 2.0~\cite{singh2022scirepeval}, and Dragon~\cite{lin2023train}.

In addition, we include biomedical-specific dense retrievers that are pretrained or fine-tuned on biomedical-specific corpora. These include MedCPT~\cite{jin2023medcpt}, BMRetriever~\cite{xu2024bmretriever}, and SPECTER 2.0, which is adapted for scientific literature retrieval. These models aim to better capture the specialized terminology and semantics of biomedical texts.

\paragraph{\textbf{Query Expansion.}}
We implement a classical biomedical query expansion method based on MetaMap~\cite{aronson2010metamap}, extracting concepts from queries restricted to semantic types: \texttt{dsyn}, \texttt{sosy}, \texttt{neop}, \texttt{patf}, and \texttt{fndg} \cite{di2019exploring}. Definitions for each concept are retrieved from the UMLS Metathesaurus using \texttt{SNOMEDCT\_US}, \texttt{MeSH}, \texttt{NCI}, and \texttt{CSP} vocabularies. We repeat the original query ten times to assign it higher weight, following the weighting strategy used in the paper.

For \textit{Query2Doc}, we follow the configuration from the original paper~\cite{wang2023query2doc}, using four in-context examples, a temperature of 1.0, and a 128-token generation limit. We use GPT-4o for generation to ensure a fair comparison with our method, which uses the same model. The pseudo-document is concatenated with the original query using $\alpha = 5$, and retrieval is performed using BM25. We also report results from the Corpus-Steered Query Expansion (CSQE) paper~\cite{lei-etal-2024-csqe} for comparison.

{\scriptsize
\renewcommand{\arraystretch}{1.4} 
\begin{table}[ht]
    \caption{IR performance (NDCG@10) across datasets}
    \label{tab:ndcg_comparison}
    \centering
    \begin{adjustbox}{max width=\textwidth}
    \begin{tabular}{
        | >{\hspace{4pt}}l<{\hspace{4pt}} 
        | >{\hspace{4pt}}c<{\hspace{4pt}} 
        | >{\hspace{4pt}}c<{\hspace{4pt}} 
        | >{\hspace{4pt}}c<{\hspace{4pt}} |
    }
        \hline
        \multirow{2}{*}{\textbf{Method}} 
        & \multicolumn{3}{c|}{\textbf{Dataset}} \\
        \cline{2-4}
        & \textbf{NFCorpus} 
        & \textbf{TREC-COVID} 
        & \textbf{SciFact} \\
        \hline
        BM25\tablefootnote{\label{fn:bm25}We use the Pyserini BM25 implementation. URL: \url{https://github.com/castorini/pyserini}}             &   0.325    &   0.656    &   0.665   \\
        \hline
        Metamap Definitions \cite{di2019exploring}   &   0.330    &  0.637   &  0.667     \\
        \hline
        Query2Doc \cite{wang2023query2doc}      & 0.303      &  0.667     & \underline{\textbf{0.711}}      \\
        \hline
        CSQE  \cite{lei-etal-2024-csqe}     & -      &  0.742     &  0.696      \\
        \hline
        Contriever \cite{izacard2021unsupervised}      &   0.328    &   0.596    &  0.677     \\
        \hline
        Dragon  \cite{lin2023train}     &  0.339     &  \underline{\textbf{0.759}}     & 0.679      \\
        \hline
        SPECTER 2.0 \cite{singh2022scirepeval}     &   0.228    &    0.584   &   0.671    \\

        \hline
        SGPT-1.3B \cite{muennighoff2022sgpt}        &   0.320    &   0.730    &  0.682     \\
        \hline
        MedCPT  \cite{jin2023medcpt}       &   \underline{\textbf{0.340}}    &  0.697     &  \underline{\textbf{0.724}}     \\
        \hline
        GTR-L \cite{ni2021large}         &  0.329     &  0.557     & 0.639      \\
        \hline
        InstructOR-L \cite{su2022one}    &   \underline{\textbf{0.341}}    &  0.581     &  0.643     \\
        \hline
        BMRetriever-410M \cite{xu2024bmretriever}      &    0.321   &   \underline{\textbf{0.831}}    &  \underline{\textbf{0.711}}     \\
        \hline
        \textit{BMQExpander} (Ours)  &   \underline{\textbf{0.363}}    &  \underline{\textbf{0.801}}     & 0.704      \\
        \hline
    \end{tabular}
    \end{adjustbox}
\end{table}
}

\mycomment{
Specifically, we use MetaMap to extract medical concepts from queries, restricting to selected semantic types to ensure relevance (\texttt{dsyn} (Disease or Syndrome), \texttt{sosy} (Sign or Symptom), etc.). For each extracted concept, we retrieve its definition from the UMLS Metathesaurus, constraining the selection to definitions provided by the SNOMED CT (\texttt{SNOMEDCT\_US}), MeSH (\texttt{MSH}), National Cancer Institute Thesaurus (\texttt{NCI}), and CRISP Thesaurus (\texttt{CSP}) vocabularies.
The expanded query is formed by repeating the original query with a weight of $\alpha = 10$ and concatenating the retrieved definitions with a weight of $1$, maintaining a strong emphasis on the original user intent similar to \cite{di2019exploring}. Retrieval is performed using BM25.
}

\subsection{Existing Datasets Experiment}

Table~\ref{tab:ndcg_comparison} presents the NDCG@10 scores of our method compared to a range of baselines, including traditional sparse retrievers, query expansion methods, and recent dense retrievers. 

The top 3 results for each dataset is highlighted. Our method achieves the highest NDCG@10 on \textbf{NFCorpus}, outperforming strong baselines such as MedCPT  and InstructOR-L. On \textbf{TREC-COVID}, it reaches 0.801, closely trailing BMRetriever-410M score of 0.831 and outperforming all the other methods. For \textbf{SciFact}, we report 0.704, which is competitive with other strong baselines. The resulting improvements over BM25 are substantial - +11.7\%, +22.1\%, and +5.9\% relative improvement on \textsc{NFCorpus}, \textsc{TREC-COVID}, and \textsc{SciFact}, respectively - showing that BMQExpander can close much of the gap to strong dense retrievers while preserving BM25's efficiency. SciFact contains queries with general academic language and fewer biomedical terms, limiting the impact of our ontology-based expansion and resulting in smaller gains.

These results demonstrate that our method performs consistently well across datasets with different biomedical content and query styles.

\begin{table}[ht]
\caption{Ablation experiment}
\label{tab:ablation_single}
\centering
\scriptsize
\setlength{\tabcolsep}{10pt}
\renewcommand{\arraystretch}{1.8}
\begin{tabular}{|l||c|c|c|}
\hline
\textbf{Configuration} & \textbf{NDCG@10} & \textbf{mAP@10} & \textbf{Recall@10} \\
\cdashline{1-4}[2pt/1pt]
\textbf{BM25 Baseline} & 0.656 & 0.016 & 0.018 \\
\cdashline{1-4}[2pt/1pt]
w/o LLM & 0.701 & 0.018 & 0.019 \\
Definitions Only & 0.750 & 0.019 & 0.021 \\
Relations Only & 0.779 & 0.020 & 0.022 \\
Definitions + Relations & \underline{\boldmath 0.801} & \underline{\boldmath 0.021} & \underline{\boldmath 0.023} \\
\hline
\end{tabular}
\end{table}

\begin{table*}[ht]
\caption{IR performance under query perturbation}
\label{tab:perturbed_methods}
\centering
\scriptsize
\begin{adjustbox}{max width=\textwidth}
\begin{tabular}{|l||ccc||ccc||ccc|}
\hline
\multirow{2}{*}{\textbf{Method}} 
& \multicolumn{3}{c||}{\textbf{NFCorpus-P}} 
& \multicolumn{3}{c||}{\textbf{TREC-COVID-P}} 
& \multicolumn{3}{c|}{\textbf{SciFact-P}} \\
\cline{2-10}
& \textbf{NDCG@10} & \textbf{mAP@10} & \textbf{Recall@10}
& \textbf{NDCG@10} & \textbf{mAP@10} & \textbf{Recall@10}
& \textbf{NDCG@10} & \textbf{mAP@10} & \textbf{Recall@10} \\
\hline
BMRet-410     & 0.245 & 0.083 & 0.119 & 0.606 & 0.016 & 0.018 & 0.677 & 0.632 & 0.801 \\
\hline
MedCPT        & 0.272 & 0.092 & 0.130 & 0.445 & 0.010 & 0.013 & 0.661 & 0.606 & 0.815 \\
\hline
Contriever    & 0.288 & 0.101 & 0.137 & 0.453 & 0.010 & 0.013 & 0.675 & 0.626 & 0.812 \\
\hline
InstructOR    & 0.291 & 0.096 & 0.139 & 0.437     & 0.009 & 0.011 & 0.623 & 0.576 & 0.752 \\
\hline
Query2Doc     & 0.307       & 0.113 & 0.153 & 0.613       & 0.016 & 0.017 & 0.701       & 0.651 & 0.837 \\
\hline
\textit{BMQExpander} (Ours) & \textbf{0.342} & \textbf{0.126} & \textbf{0.171} & \textbf{0.709} & \textbf{0.018} & \textbf{0.020} & \textbf{0.705} & \textbf{0.655} & \textbf{0.842} \\
\hline
\end{tabular}
\end{adjustbox}
\end{table*}

\begin{table*}[ht]
\caption{Comparative IR performance for different LLM architectures and prompting strategies}
\label{tab:metric_comparison}
\centering
\scriptsize
\begin{adjustbox}{max width=\textwidth}
\setlength{\tabcolsep}{6pt}  
\renewcommand{\arraystretch}{1.3}  
\begin{tabular}{|l|l||c|c|c||c|c|c||c|c|c|}
\hline
\multirow{2}{*}{\textbf{Model}} & \multirow{2}{*}{\textbf{Prompting}} 
& \multicolumn{3}{c||}{\textbf{NFCorpus}} 
& \multicolumn{3}{c||}{\textbf{TREC-COVID}} 
& \multicolumn{3}{c|}{\textbf{SciFact}} \\
\cline{3-11}
 & & \textbf{NDCG@10} & \textbf{mAP@10} & \textbf{Recall@10} 
   & \textbf{NDCG@10} & \textbf{mAP@10} & \textbf{Recall@10} 
   & \textbf{NDCG@10} & \textbf{mAP@10} & \textbf{Recall@10} \\
\cdashline{1-11}[2pt/1pt]  
\multicolumn{2}{|c||}{\textbf{BM25 Baseline}} 
  & 0.325 & 0.124 & 0.157 & 0.656 & 0.016 & 0.018 & 0.665 & 0.616 & 0.799 \\
\cdashline{1-11}[2pt/1pt]
\multirow{2}{*}{LLaMA 3.1 8B} 
  & Zero Shot & 0.353 & 0.138 & 0.182 & 0.734 & 0.019 & 0.021 & 0.695 & 0.642 & 0.844  \\
\cline{2-11}
  & CoT       & 0.359 & 0.140 & 0.183 & 0.763 & 0.020 & 0.022 & 0.695 & 0.646 & 0.826 \\
\hline
\multirow{2}{*}{Qwen 2.5 32B} 
  & Zero Shot & 0.360 & 0.141 & 0.178 & 0.748 & 0.019 & 0.021 & 0.701 & 0.653 & 0.830  \\
\cline{2-11}
  & CoT       & 0.358 & 0.136  & 0.179 & 0.760 & 0.019 & 0.021 & \underline{\boldmath 0.707} & 0.658 & 0.837 \\
\hline
\multirow{2}{*}{DeepSeek-V3} 
  & Zero Shot & \underline{\boldmath 0.369} & 0.144 & 0.188  & 0.781 & 0.020  & 0.022  & 0.716 & 0.666 & 0.854 \\
\cline{2-11}
  & CoT       & 0.365 & 0.142 & 0.184 & \underline{\boldmath 0.800}  & 0.021 & 0.022 & \underline{\boldmath 0.723} & 0.675 & 0.857 \\
\hline
\multirow{2}{*}{Gemini 1.5 Pro} 
  & Zero Shot & 0.360 & 0.140 & 0.181 & 0.764 & 0.020 & 0.022 & 0.705 & 0.659 & 0.832 \\
\cline{2-11}
  & CoT       & 0.361 & 0.140 & 0.181 & 0.771 & 0.020 & 0.028 & 0.695 & 0.646 & 0.829 \\
\hline

\multirow{2}{*}{GPT-4 Omni} 
  & Zero Shot & \underline{\boldmath 0.365} & 0.143 & 0.187 & 0.785  & 0.020 & 0.022 & 0.700 & 0.649 & 0.838 \\
\cline{2-11}
  & CoT       & 0.363  & 0.141 & 0.181 & \underline{\boldmath 0.801}  & 0.021 & 0.023 & 0.704 & 0.654 & 0.845 \\
\hline

\end{tabular}
\end{adjustbox}
\end{table*}

This ablation study evaluates the contribution of each component in \textit{BMQExpander} on the TREC-COVID dataset. Using only ontology-derived keywords - omitting any LLM expansion generation - raises NDCG@10 from 0.656 to 0.701. Adding either definitions or relations further improves performance to 0.750 and 0.779, showing that each adds useful context. The full method - combining definitions, relations, and the LLM pseudo-document achieves the best results, highlighting the importance of integrating all components for optimal retrieval.

\subsection{Query Perturbation Experiment}

Table~\ref{tab:perturbed_methods} reports retrieval performance on paraphrased queries across three benchmark datasets. All dense retrievers exhibit notable performance degradation especially on TREC-COVID-P and NFCorpus-P datasets. The decline in performance can be attributed to dense retrievers’ sensitivity to distribution shifts, as paraphrased queries can alter embedding representations in ways that disrupt learned query-document alignment.

Query expansion methods demonstrate greater resilience to query perturbation, showing only moderate performance drops, while our method consistently outperforms all baselines by a substantial margin.

These results highlight BMQExpander’s robustness and effectiveness under semantic variations in user queries.


\subsection{Experiments on LLM Architectures and Prompting Methods}

Table~\ref{tab:metric_comparison} presents a comparative analysis of retrieval performance across various large language models (LLMs) under two prompting strategies: zero-shot and chain-of-thought (CoT). The evaluation includes open-source LLaMA 3.1 (8B), DeepSeek-V3, Qwen 2.5 (32B), and commercially available GPT-4 Omni and Gemini 1.5 Pro.

All LLM variants - regardless of size or complexity - deliver large gains over the BM25 baseline, confirming that our generative expansion is model‐agnostic. DeepSeek-V3 attains the best NDCG@10 score of 0.369 and 0.723 on \textsc{NFCorpus} and \textsc{SciFact} respectively, while GPT-4o leads on \textsc{TREC-COVID} with 0.801.

CoT prompting is most beneficial on the \textsc{TREC-COVID} set, but its impact is uneven across model sizes. Smaller models like LLaMA shows the largest relative lift of 4\% in NDCG.  In contrast, mid and large-scale models (Qwen, DeepSeek-V3, GPT-4o, Gemini 1.5) record only modest increases of 0.7–1.9 NDCG points, and CoT brings little or no benefit on \textsc{NFCorpus} and \textsc{SciFact}.

Overall, the results show that \textit{BMQExpander} is broadly transferable across LLM backbones: performance differences among the tested models remain modest, and although CoT reasoning noticeably boosts smaller architectures, its incremental value is limited when averaged across datasets.

\subsection{Qualitative Clinical Evaluation}

Following established methodologies for evaluating medical language model outputs \cite{singhal2025toward, zuo2024medhallbench, shool2025systematic}, we conducted a controlled expert evaluation to assess the clinical fidelity of pseudo-documents generated by our ontology-guided approach compared to two other LLM-based query expansion methods.

\paragraph{Study design.}
We randomly sampled 15 queries from three biomedical benchmarks and generated one pseudo-document per query with each of the three systems—\textsc{BMQExpander}, Query2Doc, and CSQE.
To eliminate model‐specific variance, all generations were produced with the same LLM backbone, \textit{GPT-3.5-turbo}, matching the configuration originally adopted by CSQE.
For every query, the three outputs were independently shuffled and evaluated under single-blind conditions by an experienced, board-certified medical professional.
The evaluation yielded a total of 135 scalar judgments (15 queries × 3 methods × 3 rating criteria).

\paragraph{Rating criteria.}
To capture the main risks and benefits of long-form medical expansions, we scored every pseudo-document on three criteria:

\vspace{5pt}

 \textbf{Medical Accuracy} – Evaluates the factual correctness of biomedical claims against peer-reviewed literature and established clinical guidelines \cite{singhal2025toward, shool2025systematic} on a five-point Likert scale from substantial inaccuracies to complete verifiability.
Five-point scale:
\emph{1} = contains multiple factual errors that contradict established medical knowledge,
\emph{2} = contains several minor factual inaccuracies,
\emph{3} = mostly accurate with occasional imprecision in terminology or details,
\emph{4} = accurate with negligible imprecision,
\emph{5} = all medical claims are factually correct and verifiable.

\vspace{5pt}
 \textbf{Query Relevance} – Measures how directly and completely the expansion answers the user’s question \cite{shool2025systematic, ju2023improving} on a five-point Likert scale from irrelevant output to comprehensive query coverage.  

      Five-point scale:  
      \emph{1} = content is largely unrelated to the query focus,  
      \emph{2} = partially addresses the query but misses key aspects,
      \emph{3} = covers main query elements but lacks depth or completeness,  
      \emph{4} = thoroughly addresses most query components,
      \emph{5} = provides comprehensive, targeted response covering all relevant aspects of the clinical question.

\vspace{5pt}
 \textbf{Hallucination Severity} – Assesses the presence and clinical risk of fabricated, unverifiable, or speculative medical content \cite{asgari2025framework} on a three-point scale from no fabrications to major fabrications risking clinical decision-making.  

      Three-level scale:  
      \emph{0} = no fabricated content detected,  
      \emph{1} = minor fabrications present but clinically inconsequential (e.g., imprecise statistics, overgeneralized statements),  
      \emph{2} = major fabrications that could mislead clinical decision-making or patient care (e.g., false contraindications, invented treatment protocols).



\begin{table}[ht]
  \centering
  \scriptsize           
  \caption{Expert evaluation of clinical quality}
  \label{tab:clinical}
  \begin{tabular}{lccc}
    \toprule
    \textbf{Method} &
    \textbf{Accuracy} $\uparrow$ &
    \textbf{Relevance} $\uparrow$ &
    \textbf{Hallucination} $\downarrow$ \\ \midrule
    \textsc{BMQExpander} & $4.47 \pm 0.92$ & $4.40 \pm 1.06$ & $0.07 \pm 0.26$ \\
    Query2Doc            & $4.00 \pm 1.13$ & $3.13 \pm 1.55$ & $0.07 \pm 0.26$ \\
    CSQE                 & $3.53 \pm 0.74$ & $4.00 \pm 1.41$ & $0.80 \pm 0.56$ \\ \bottomrule
  \end{tabular}
  
  \vspace{5pt}      
  \footnotesize
  Higher values indicate better \textbf{Accuracy} and \textbf{Relevance};
  lower values indicate fewer \textbf{Hallucinations}.
  All numbers are reported as $\mu\!\pm\!\sigma$.
\end{table}

\paragraph{Results.}
Table \ref{tab:clinical} summarizes the mean $\pm$ s.d.\ scores.
\textsc{BMQExpander} attains the highest accuracy and relevance.
Crucially, ontology grounding limits hallucination by an order of magnitude lower than CSQE and on par with Query2Doc.

We understand that a larger qualitative study is needed, which is challenging given the availability of medical experts. We leave this as future work.


\section{Conclusion}

This work introduced BMQExpander, an ontology-guided framework that embeds UMLS definitions and carefully pruned relations into large-language-model prompts to perform query expansion. Across three biomedical benchmarks, BMQExpander consistently outperformed strong sparse, dense, and LLM-based query expansion baselines; ablations confirmed that definitions and relations each add measurable value, and the improvements persisted across five different LLM backbones.
Experiments on perturbed datasets show that BMQExpander generalizes much better than state-of-the-art biomedical retrieval methods.








\printbibliography


\end{document}